\documentclass{ita}
\usepackage[latin1]{inputenc}

\usepackage{amssymb}

\def\ufootnote#1{\let\savedthfn\thefootnote\let\thefootnote\relax
\footnote{#1}\let\thefootnote\savedthfn\addtocounter{footnote}{-1}}

\newcommand{\fa}{\forall}
\newcommand{\Ga}{\Gamma}
\newcommand{\Gas}{\Gamma^\star}
\newcommand{\Gao}{\Gamma^\omega}
\newcommand{\Si}{\Sigma}
\newcommand{\Sis}{\Sigma^\star}
\newcommand{\Sio}{\Sigma^\omega}
\newcommand{\ra}{\rightarrow}
\newcommand{\hs}{\hspace{12mm}

\noi}
\newcommand{\lra}{\leftrightarrow}

\newcommand{\ol}{ $\omega$-language}
\newcommand{\orl}{ $\omega$-regular language}
\newcommand{\om}{\omega}
\newcommand{\nl}{\newline}
\newcommand{\noi}{\noindent}

\newcommand{\trans}[1]{\stackrel{#1}{\rightarrow}}
\newcommand{\dom}{\operatorname{Dom}}
\newcommand{\im}{\operatorname{Im}}

\begin{document}

\title{On the Continuity Set of  \\ an OMega Rational Function  
\\
\hs 
 Dedicated to Serge Grigorieff on the occasion of his 60th Birthday}

\runningtitle{ On the Continuity Set of an omega-Rational Function}

\author{Olivier Carton}
\address{LIAFA, Universit\'e Paris 7  et  CNRS  \\ 2 Place Jussieu 75251 Paris cedex 05,  France.\\
   \email{Olivier.Carton@liafa.jussieu.fr}}
\author{Olivier Finkel}
\address{Equipe Modèles de Calcul et Complexité   
 \\ \it Laboratoire de l'Informatique du Parallélisme\\ (UMR 5668 - CNRS - ENS Lyon - UCB Lyon - INRIA)
 \\  CNRS et Ecole Normale Supérieure de Lyon
 \\ 46, Allée d'Italie 69364 Lyon Cedex 07, France. \\ 
\email{Olivier.Finkel@ens-lyon.fr }}
\author{Pierre Simonnet}
\address{UMR 6134-Systèmes Physiques de l'Environnement \\  Facult\'e des Sciences, Universit\'e de Corse\\
 Quartier Grossetti BP52 20250, Corte, France  
\\
\email{simonnet@univ-corse.fr.} }
\date{}


\ufootnote{LIP Research Report RR 2008-04}


\begin{abstract}
  In this paper, we study the continuity of rational functions realized by
  B\"uchi finite state transducers.  It has been shown by Prieur that it
  can be decided whether such a function is continuous.  We prove here that
  surprisingly, it cannot be decided whether such a function~$f$ has
  at least one point of continuity and that its continuity set $C(f)$
  cannot be computed.

  In the case of a synchronous rational function, we show that its
  continuity set is rational and that it can be computed.  Furthermore we
  prove that any rational ${\bf \Pi}^0_2$-subset of $\Sio$ for some alphabet $\Si$
  is the continuity set $C(f)$ of an $\omega$-rational synchronous
  function~$f$ defined on $\Sio$.
\end{abstract}

\maketitle

\section{Introduction}

Acceptance of infinite words by finite automata was firstly considered in
the sixties by B\"uchi in order to study decidability of the monadic second
order theory of one successor over the integers \cite{bu62}. Then the so
called \orl s have been intensively studied and many applications have been
found.  We refer the reader to \cite{tho,sta,pp} for many results and
references.

Gire and Nivat studied infinitary rational relations accepted by B\"uchi
transducers in \cite{gire1,gn}.  Infinitary rational relations are subsets
of $\Si^\om \times \Ga^\om$, where $\Si$ and $\Ga$ are finite alphabets,
which are accepted by $2$-tape finite B\"uchi automata with two
asynchronous reading heads.  They have been much studied, in particular in
connection with the rational functions they may define, see for example
\cite{cg,bcps,sim,sta,pri}.

Gire proved in \cite{gire} that one can decide whether an infinitary
rational relation $R \subseteq \Sio \times \Gao$ recognized by a given
B\"uchi transducer $\mathcal{T}$ is the graph of a function $f : \Sio \ra
\Gao$, (respectively, is the graph of a function $f : \Sio \ra \Gao$
recognized by a synchronous B\"uchi transducer). Such a function is called
an $\om$-rational function (respectively, a synchronous $\om$-rational
function).

The continuity of $\om$-rational functions is an important issue since it
is related to many aspects.  Let us mention two of them.  First,
sequential functions that may be realized by input deterministic automata
are continuous but the converse is not true.  Second, continuous functions
define a reduction between subsets of a topological space that yields a hierarchy
called the Wadge hierarchy.  The restriction of this hierarchy to rational
sets gives the Wagner hierarchy.

This paper is focused on the continuity sets of rational functions.  Prieur
proved in \cite{pri,pri01} that it can be decided whether a given
$\om$-rational function is continuous.  This means that it can be decided
whether the continuity set is equal to the domain of the function.  We show
however that it cannot be decided whether a rational function has at least one
point of continuity.  We show that in general the continuity set of a
rational function is not rational and even not context-free.  Furthermore,
we prove that it cannot be decided whether this continuity set is rational.

We pursue this study with synchronous rational functions.  These functions
are accepted by B\"uchi transducers in which the two heads move
synchronously.  Contrary to the general case, the continuity set of
synchronous rational function is always rational and it can be effectively
computed.  We also give a characterization of continuity sets of
synchronous functions.  It is well known that a continuity set is a
$\mathbf{\Pi}^0_2$-set.  We prove conversely that any rational
$\mathbf{\Pi}^0_2$-set is the continuity set of some synchronous rational
function.

The paper is organized as follows. In section 2 we recall the notions of
infinitary rational relation, of $\omega$-rational function, of synchronous
or asynchronous $\omega$-rational function, of topology and continuity; we
recall also some recent results on the topological complexity of infinitary
rational relations.  In section 3 we study the continuity sets of
$\omega$-rational functions in the general case, stating some
undecidability results.  Finally we study the case of synchronous
$\omega$-rational functions in section~4.

\section{Recall  on $\om$-rational functions and topology}
\subsection{Infinitary rational relations and $\om$-rational functions}

Let $\Si$ be a finite alphabet whose elements are called letters.  A
non-empty finite word over $\Si$ is a finite sequence of letters:
$x=a_1a_2\ldots a_n$ where $\fa i\in [1; n]$ $a_i \in\Si$.  We shall denote
$x(i)=a_i$ the $i^{th}$ letter of $x$ and $x[i]=x(1)\ldots x(i)$ for $i\leq
n$. The length of $x$ is $|x|=n$.  The empty word will be denoted by
$\lambda$ and has 0 letter. Its length is 0.  The set of finite words over
$\Si$ is denoted $\Sis$.  $\Si^+ = \Sis - \{\lambda\}$ is the set of non
empty words over $\Si$.  A (finitary) language $L$ over $\Si$ is a subset
of $\Sis$.  The usual concatenation product of $u$ and $v$ will be denoted
by $u.v$ or just $uv$.  For $V\subseteq \Sis$, we denote by $V^\star$ the
set $R=\{ v_1\ldots v_n \mid n\in \mathbb{N} \text{ and } \fa i \in [1; n]
\;\; v_i \in V \}$.

The first infinite ordinal is $\om$.  An $\om$-word over $\Si$ is an $\om$
-sequence $a_1a_2\ldots a_n \ldots$, where for all integers $i\geq 1$ $a_i
\in\Sigma$.  When $\sigma$ is an $\om$-word over $\Si$, we write $\sigma
=\sigma(1)\sigma(2)\ldots \sigma(n) \ldots $ and
$\sigma[n]=\sigma(1)\sigma(2)\ldots \sigma(n)$ the finite word of length
$n$, prefix of $\sigma$.  The set of $\om$-words over the alphabet $\Si$ is
denoted by $\Si^\om$.  An $\om$-language over an alphabet $\Sigma$ is a
subset of $\Si^\om$.  For $V\subseteq \Sis$, $V^\om = \{ \sigma =u_1\ldots
u_n\ldots \in \Si^\om \mid \fa i\geq 1 ~~ u_i\in V \}$ is the $\om$-power
of $V$.  The concatenation product is extended to the product of a finite
word $u$ and an $\om$-word $v$: the infinite word $u.v$ is then the
$\om$-word such that: $(u.v)(k)=u(k)$ if $k\leq |u|$ , and
$(u.v)(k)=v(k-|u|)$ if $k>|u|$.

We assume the reader to be familiar with the theory of formal languages and
of \orl s, see \cite{bu62,tho,eh,sta,pp} for many results and references.
We recall that \orl s form the class of \ol s accepted by finite automata
with a B\"uchi acceptance condition and this class, denoted by $RAT$,  is the omega
Kleene closure of the class of regular finitary languages.

We are going now to recall the notion of infinitary rational relation which
extends the notion of \orl , via definition by B\"uchi transducers:

\begin{dfntn}
  A B\"uchi transducer is a sextuple $\mathcal{T}=(K, \Si, \Ga, \Delta,
  q_0, F)$, where $K$ is a finite set of states, $\Si$ and $\Ga$ are finite
  sets called the input and the output alphabets, $\Delta$ is a finite
  subset of $K \times \Sis \times \Gas \times K$ called the set of
  transitions, $q_0$ is the initial state, and $F \subseteq K$ is the set
  of accepting states.  \nl A computation $\mathcal{C}$ of the transducer
  $\mathcal{T}$ is an infinite sequence of consecutive transitions
  \begin{displaymath}
  (q_0, u_1, v_1, q_1), (q_1, u_2, v_2, q_2), \ldots 
  (q_{i-1}, u_{i}, v_{i}, q_{i}),  (q_i, u_{i+1}, v_{i+1}, q_{i+1}), \ldots 
  \end{displaymath}
  The computation is said to be successful iff there exists a final state
  $q_f \in F$ and infinitely many integers $i\geq 0$ such that $q_i=q_f$.
  The input word and output word of the computation are respectively
  $u=u_1.u_2.u_3 \ldots$ and $v=v_1.v_2.v_3 \ldots$ The input and the
  output words may be finite or infinite.  The infinitary rational relation
  $R(\mathcal{T})\subseteq \Sio \times \Ga^\om$ accepted by the B\"uchi
  transducer $\mathcal{T}$ is the set of couples $(u, v) \in \Sio \times
  \Ga^\om$ such that $u$ and $v$ are the input and the output words of some
  successful computation $\mathcal{C}$ of $\mathcal{T}$.  The set of
  infinitary rational relations will be denoted $RAT_2$.
\end{dfntn} 

If $R(\mathcal{T})\subseteq \Sio \times \Ga^\om$ is an infinitary rational
relation recognized by the B\"uchi transducer $\mathcal{T}$ then we denote
$\dom(R(\mathcal{T}))=\{ u \in \Sio \mid \exists v \in \Gao ~~(u, v) \in
R(\mathcal{T}) \}$ and $\im(R(\mathcal{T}))=\{ v \in \Gao \mid \exists u \in
\Sio (u, v) \in R(\mathcal{T}) \}$.

It is well known that, for each infinitary rational relation
$R(\mathcal{T})\subseteq \Sio \times \Ga^\om$, the sets
$\dom(R(\mathcal{T}))$ and $\im(R(\mathcal{T}))$ are regular $\om$-languages.

The B\"uchi transducer $\mathcal{T}=(K, \Si, \Ga, \Delta, q_0, F)$ is said
to be synchronous if the set of transitions $\Delta$ is a finite subset of
$K \times \Si \times \Ga \times K$, i.e. if each transition is labelled
with a pair $(a, b)\in \Si \times \Ga$.  An infinitary rational relation
recognized by a synchronous B\"uchi transducer is in fact an $\om$-language
over the product alphabet $\Si \times \Ga$ which is accepted by a B\"uchi
automaton. It is called a synchronous infinitary rational relation. An
infinitary rational relation is said to be asynchronous if it can not be
recognized by any synchronous B\"uchi transducer.  Recall now the following
undecidability result of C. Frougny and J. Sakarovitch.

\begin{thrm}[\cite{fs}]
  One cannot decide whether a given infinitary rational relation is
  synchronous.
\end{thrm}

A B\"uchi transducer $\mathcal{T}=(K, \Si, \Ga, \Delta, q_0, F)$ is said to
be functional if for each $u\in \dom(R(\mathcal{T}))$ there is a unique
$v\in \im(R(\mathcal{T}))$ such that $(u, v) \in R(\mathcal{T})$.  The
infinitary rational relation recognized by $\mathcal{T}$ is then a
functional relation and it defines an $\om$-rational (partial) function
$f_{\mathcal{T}}: \dom(R(\mathcal{T})) \subseteq \Sio \ra \Gao$ by:
\newline for each $u\in \dom(R(\mathcal{T}))$, $f_{\mathcal{T}}(u)$ is the unique
$v\in\Gao$ such that $(u, v) \in R(\mathcal{T})$.  \nl An $\om$-rational
(partial) function $f: \Sio \ra \Gao$ is said to be synchronous if there is
a synchronous B\"uchi transducer $\mathcal{T}$ such that
$f=f_{\mathcal{T}}$.  \nl An $\om$-rational (partial) function $f: \Sio \ra
\Gao$ is said to be asynchronous if there is no synchronous B\"uchi
transducer $\mathcal{T}$ such that $f=f_{\mathcal{T}}$.

\begin{thrm}[\cite{gire}]
  One can decide whether an infinitary rational relation recognized by a
  given B\"uchi transducer $\mathcal{T}$ is a functional infinitary
  rational relation (respectively, a synchronous functional infinitary
  rational relation).
\end{thrm}

\subsection{Topology}

We assume the reader to be familiar with basic notions of topology which
may be found in \cite{mos,kec,lt,sta,pp}.  There is a natural metric on
the set $\Sio$ of infinite words over a finite alphabet $\Si$ which is
called the prefix metric and defined as follows. For $u, v \in \Sio$ and
$u\neq v$ let $d(u, v)=2^{-l_{pref(u,v)}}$ where $l_{pref(u,v)}$ is the
least integer $n$ such that the $(n+1)^{th}$ letter of $u$ is different
from the $(n+1)^{th}$ letter of $v$.  This metric induces on $\Sio$ the
usual Cantor topology for which open subsets of $\Sio$ are in the form
$W.\Si^\om$, where $W\subseteq \Sis$.  Recall that a set $L\subseteq
\Si^\om$ is a closed set iff its complement $\Si^\om - L$ is an open set.
We define now the next classes of the Borel Hierarchy:

\begin{dfntn}
The classes ${\bf \Si}_n^0$ and ${\bf \Pi}_n^0 $ of the Borel Hierarchy
 on the topological space $\Si^\om$  are defined as follows:
 \begin{itemize}
  \item[-] ${\bf \Si}^0_1 $ is the class of open sets of $\Si^\om$.
  \item[-] ${\bf \Pi}^0_1 $ is the class of closed sets of $\Si^\om$.
  \end{itemize}
  And for any integer $n\geq 1$:
  \begin{itemize}
  \item[-] ${\bf \Si}^0_{n+1}$  is the class of countable unions of ${\bf
      \Pi}^0_n $-subsets of $\Si^\om$.
  \item[-] ${\bf \Pi}^0_{n+1}$ is the class of countable intersections of
    ${\bf \Si}^0_n$-subsets of $\Si^\om$.
  \end{itemize}
The Borel Hierarchy is also defined for transfinite levels: 
The classes ${\bf \Si}^0_\alpha $ and ${\bf \Pi}^0_\alpha $,
  for a non-null countable ordinal $\alpha$, are defined in the following
  way.
  \begin{itemize}
  \item[-] ${\bf \Si}^0_\alpha $ is the class of countable unions of
    subsets of $\Si^\om$ in $\cup_{\gamma <\alpha}{\bf \Pi}^0_\gamma $.
  \item[-] ${\bf \Pi}^0_\alpha $ is the class of countable intersections of
    subsets of $\Si^\om$ in $\cup_{\gamma <\alpha}{\bf \Si}^0_\gamma $.
  \end{itemize}
\end{dfntn}

Let us recall the characterization of rational ${\bf \Pi}^0_2 $-subsets of $\Sio$,
due to Landweber \cite{la}.  This characterization will  be used in the
proof that any rational ${\bf \Pi}^0_2$-subset is the continuity set of
some rational synchronous function.

\begin{thrm}[Landweber] \label{thm:landweber}
  A rational subset of $\Sigma^\omega$ is ${\bf \Pi}^0_2 $ if and only if
  it can be recognized by a deterministic B\"uchi automaton.
\end{thrm}

There are some subsets of the Cantor set, (hence also of the topological
space $\Sio$, for a finite alphabet $\Si$ having at least two elements)
which are not Borel sets. There exists another hierarchy beyond the Borel
hierarchy, called the projective hierarchy.  The first class of the
projective hierarchy is the class ${\bf \Si}^1_1$ of {\bf analytic} sets.
A set $A \subseteq \Sio$ is analytic iff there exists a Borel set $B
\subseteq (\Si \times Y)^\om$, with $Y$ a finite alphabet, such that $ x
\in A \lra \exists y \in Y^\om $ such that $(x, y) \in B$, where $(x, y)\in
(\Si \times Y)^\om$ is defined by: $(x, y)(i)=(x(i),y(i))$ for all integers
$i\geq 1$.  \nl

\begin{rmrk}
An infinitary rational relation is a subset of $\Sio \times \Ga^\om$ for two finite 
alphabets $\Si$ and $\Ga$. One can also consider that it is an \ol~ over the finite 
alphabet $\Si \times \Ga$. If $(u, v) \in \Sio \times \Ga^\om$,
 one can consider this couple of 
infinite words as a single infinite word $(u(1),v(1)).(u(2), v(2)).(u(3), v(3))\ldots $
over the alphabet $\Si \times \Ga$. 
 Since the set $(\Si \times\Ga)^\omega$ of infinite words over the  finite alphabet $\Si \times \Ga$ is naturally equipped with the Cantor topology, it is natural to 
 investigate the topological complexity 
of infinitary rational relations as $\om$-languages, and to locate them with regard to the Borel and projective hierarchies.  
Every infinitary rational relation is an analytic set and there exist some ${\bf \Si}^1_1$-complete, hence non-Borel, infinitary rational relations \cite{relrat}.
The second author has recently proved the following very surprising result: 
 infinitary rational relations have the  same 
topological complexity as $\om$-languages accepted by B\"uchi Turing machines \cite{mscs06,stacs06}. In particular, 
for every recursive non-null ordinal 
$\alpha$ there exist some ${\bf \Pi}^0_\alpha$-complete and some ${\bf \Si}^0_\alpha$-complete infinitary rational relations, 
and the supremum of the set of Borel ranks of  infinitary rational relations is   the ordinal $\gamma_2^1$. 
This ordinal is defined by A.S. Kechris, D. Marker, and  R.L. Sami in 
\cite{kms} and it is proved to be strictly greater than the ordinal 
$\delta_2^1$ which is the first non $\Delta_2^1$ ordinal. Thus the ordinal  $\gamma_2^1$ is also strictly greater than the first non-recursive 
ordinal  $ \om_1^{\mathrm{CK}}$, usually called the Church-kleene ordinal. Notice that amazingly the exact value of the ordinal $\gamma_2^1$ 
may depend on axioms of set theory, see \cite{kms,mscs06}.
 \end{rmrk}

\begin{rmrk}
Infinitary rational relations  recognized by  synchronous B\"uchi transducers are regular $\om$-languages thus they are boolean combinations of 
${\bf \Pi}^0_2$-sets hence ${\bf \Delta}^0_3$-sets \cite{pp}. So we can see that there is a great difference between the cases of synchronous and of 
asynchronous infinitary rational relations. We shall see in the sequel that these two cases are also very different when we investigate the continuity sets of 
$\om$-rational functions. 
 \end{rmrk}

\subsection{Continuity}

We have already seen that the Cantor topology of a space $\Sio$ can be
defined by a distance $d$.  We recall that a function $f: \dom(f) \subseteq
\Sio \ra \Gao$, whose domain is $\dom(f)$, is said to be continuous at point
$x\in \dom(f)$ if :

$$\forall n\geq 1 ~~~\exists k \geq 1 ~~~\forall y\in \dom(f) ~~~[~ d(x, y)<2^{-k} \Rightarrow d(f(x), f(y))<2^{-n}~] $$

The function $f$ is said to be continuous if it is continuous at every
point $x\in\Sio$.

The continuity set $C(f)$ of the function $f$ is the set of points of
continuity of $f$.

Recall that if $X$ is a subset of $\Sio$, it is also a topological space whose topology is induced by the topology of $\Sio$. 
Open sets 
of $X$  are traces on $X$  of open sets of 
$\Si ^\om$ and the same result holds for closed sets. Then 
one can easily show by induction that for every integer  $n \geq 1$,  
${\bf \Pi }^0_n $-subsets 
(resp.  ${\bf \Si}^0_n  $-subsets) of  $X$  are traces on  $X$ of 
${\bf \Pi }^0_n $-subsets 
(resp.  ${\bf \Si }^0_n $-subsets) of $\Si^\om$, i.e. are 
intersections with $X$ of 
${\bf \Pi}^0_n $-subsets 
(resp.  ${\bf \Si}^0_n $-subsets) of  $\Si^\om$. 

We recall now the following well known result. 

\begin{thrm}[see \cite{kec}]
  Let $f$ be a function from $\dom(f) \subseteq \Sio $ into $\Gao$. Then the
  continuity set $C(f)$ of $f$ is always a ${\bf \Pi }^0_2 $-subset of
  $\dom(f)$.
\end{thrm}

\begin{proof}

Let $f$ be a  function from $Dom(f) \subseteq \Sio $ into $\Gao$. For some integers $n, k\geq 1$, we consider 
the set 
$$X_{k,n}=\{ x\in Dom(f) \mid \forall y\in Dom(f) ~~~[~ d(x, y)<2^{-k} \Rightarrow d(f(x), f(y))<2^{-n}~] \}$$

\noi We know, from  the definition of the distance $d$, 
that for two $\om$-words $x$ and  $y$ over $\Si$, the inequality $d(x, y)<2^{-k}$ simply means that $x$ and  $y$ 
have the same $(k+1)$ first letters. 
\nl Then it is easy to see that the set  $X_{k,n}$ is an open subset of $Dom(f)$, 
because for each $x \in X_{k,n}$, the set $X_{k,n}$ contains the open ball (in $Dom(f)$) of all 
$y\in Dom(f)$ such that $d(x, y)<2^{-k}$. 

\hs By union we can infer that $X_n=\bigcup_{k\geq 1}X_{k,n}$ is an open subset of $Dom(f)$ and then the countable intersection 
$C(f)=\bigcap_{n\geq 1}X_n$ is a ${\bf \Pi }^0_2 $-subset of $Dom(f)$. 

\end{proof}

In the sequel we are going to investigate the continuity sets of
$\om$-rational functions, firstly in the general case and next in the case
of synchronous $\om$-rational functions.

\section{Continuity set of   $\om$-rational functions}

Recall that C. Prieur proved the following result. 

\begin{thrm}[\cite{pri, pri01}]
   One can decide whether a given $\om$-rational function is continuous. 
\end{thrm}

Prieur showed that the closure (in the topological sense) of the graph of
a rational relation is still a rational relation that can be effectively
computed.  From this closure, it is quite easy to decide whether 
a given $\om$-rational function is continuous.

So one can decide whether the continuity set of an $\om$-rational function
$f$ is equal to its domain $\dom(f)$.  We shall prove below some
undecidability results, using the undecidability of the \emph{Post
  Correspondence Problem} which we now recall.

\begin{thrm}[Post]
  Let $\Ga $ be an alphabet having at least two elements . Then it is
  undecidable to determine, for arbitrary n-tuples $(u_1,\ldots ,u_n)$ and
  $(v_1,\ldots ,v_n)$ of non-empty words in $\Gas$, whether there exists a
  non-empty sequence of indices $i_1,\ldots ,i_k $ such that $u_{i_1}\ldots
  u_{i_k} =v_{i_1}\ldots v_{i_k}$.
\end{thrm}

We now state our first   undecidability result. 

\begin{thrm}
  One cannot decide whether the continuity set $C(f)$ of a given
  $\om$-rational function $f$ is empty.
\end{thrm}

\begin{proof}
  Let $\Ga $ be an alphabet having at least two elements and $(u_1,\ldots
  ,u_n)$ and $(v_1,\ldots ,v_n)$ be two sequences of $n$ non-empty words in
  $\Gas$.  Let $A=\{a, b\}$ and $C=\{c_1, \ldots , c_n\}$ such that $A\cap
  C=\emptyset$ and $A\cap \Ga=\emptyset$.

  We define the $\om$-rational function $f$ of domain $\dom(f)=C^+.A^\om$
  by:
  \begin{itemize}
  \item $f(x)=u_{i_1}\ldots u_{i_k}.z $~~if $x=c_{i_1}\ldots c_{i_k}.z$ and
    $z\in (A^\star.a)^\om$.  
  \item $f(x)=v_{i_1}\ldots v_{i_k}.z $~~if $x=c_{i_1}\ldots c_{i_k}.z$ and
    $z\in A^\star.b^\om$. 
  \end{itemize}

  Notice that $(A^\star.a)^\om$ is simply the set of $\om$-words over the alphabet $A$ having an infinite number of occurrences of the letter 
  $a$. And $A^\star.b^\om$ is the complement of $ (A^\star.a)^\om$ in $A^\om$, i.e. it is the set  of $\om$-words over the alphabet $A$ containing only 
  a finite number of letters $a$. 

  The two $\om$-languages $(A^\star.a)^\om$ and $A^\star.b^\om$ are
  $\om$-regular, so they are accepted by B\"uchi automata.  It is then easy
  to see that the function $f$ is $\om$-rational and we can construct a
  B\"uchi transducer $\mathcal{T}$ that accepts the graph of~$f$.

  We are going to prove firstly that if $x=c_{i_1}\ldots c_{i_k}.z \in
  C^+.A^\om$ is a point of continuity of the function $f$ then the Post
  Correspondence Problem of instances $(u_1,\ldots ,u_n)$ and $(v_1,\ldots
  ,v_n)$ would have a solution $i_1,\ldots ,i_k $ such that $u_{i_1}\ldots
  u_{i_k} =v_{i_1}\ldots v_{i_k}$.

  We distinguish two cases. 

  \paragraph{First Case.} Assume firstly that $z\in (A^\star.a)^\om$. 
  Then by definition of $f$ it holds that $f(x)=u_{i_1}\ldots u_{i_k}.z $. 
  Notice that there is a sequence of elements $z_n \in  A^\star.b^\om$, $n\geq 1$, such that  the sequence $(z_n)_{n\geq 1}$ is convergent and 
  $lim (z_n) = z$. This is due to the fact that $A^\star.b^\om$ is dense in  $A^\om$. 
  We set $x_n=c_{i_1}\ldots c_{i_k}.z_n$. So we have also $lim (x_n) = x$. 
  \nl By definition of $f$, it holds that $f(x_n)=f(c_{i_1}\ldots c_{i_k}.z_n)=v_{i_1}\ldots v_{i_k}.z_n $. 
  \nl If $x=c_{i_1}\ldots c_{i_k}.z$ is a point of continuity of $f$ then we must have $lim (f(x_n))=f(x)=u_{i_1}\ldots u_{i_k}.z $.
  But $f(x_n)= v_{i_1}\ldots v_{i_k}.z_n $ converges to $v_{i_1}\ldots v_{i_k}.z $. 
  Thus  this would imply that $u_{i_1}\ldots u_{i_k}=v_{i_1}\ldots v_{i_k}$ 
  and  the Post Correspondence Problem of instances $(u_1,\ldots ,u_n)$ and $(v_1,\ldots ,v_n)$ would have  a solution. 

   \paragraph{Second  Case.} 

   Assume now that $z\in A^\star.b^\om$.  Notice that $(A^\star.a)^\om$ is
   also dense in $A^\om$. Then reasoning as in the case of $z\in
   (A^\star.a)^\om$, we can prove that if $x=c_{i_1}\ldots c_{i_k}.z$ is a
   point of continuity of $f$ then $u_{i_1}\ldots u_{i_k}=v_{i_1}\ldots
   v_{i_k}$ so the Post Correspondence Problem of instances $(u_1,\ldots
   ,u_n)$ and $(v_1,\ldots ,v_n)$ would have a solution.

   Conversely assume that the Post Correspondence Problem of instances
   $(u_1,\ldots ,u_n)$ and $(v_1,\ldots ,v_n)$ has a solution, i.e.  a
   non-empty sequence of indices $i_1,\ldots ,i_k $ such that
   $u_{i_1}\ldots u_{i_k} =v_{i_1}\ldots v_{i_k}$.  

   Consider now the
   function $f$ defined above. We have:
   $$f(c_{i_1}\ldots c_{i_k}.z)=u_{i_1}\ldots u_{i_k}.z = v_{i_1}\ldots v_{i_k}.z ~~~~\mbox{ for  every }
  z\in A^\om$$

  So it is easy to see  that the function $f$ is continuous at point $c_{i_1}\ldots c_{i_k}.z$ for every $z\in A^\om$.

  Finally we have proved that the function $f$ is continuous at point
  $c_{i_1}\ldots c_{i_k}.z$, for $z\in A^\om$, if and only if the non-empty
  sequence of indices $i_1,\ldots ,i_k $ is a solution of the Post
  Correspondence Problem of instances $(u_1,\ldots ,u_n)$ and $(v_1,\ldots
  ,v_n)$.  Thus one cannot decide whether the function $f$ has (at
  least) one point of continuity.

\end{proof}

\begin{thrm}
  One cannot decide whether the continuity set $C(f)$ of a given
  $\om$-rational function $f$ is a regular $\om$-language (respectively, a
  context-free $\om$-language).
\end{thrm}

\begin{proof}
  We shall use a particular instance of Post Correspondence Problem. For
  two letters $c, d$, let PCP$_1$ be the Post Correspondence Problem of
  instances $(t_1, t_2, t_3)$ and $(w_1, w_2, w_3)$, where $t_1=c^2$,
  $t_2=t_3=d$ and $w_1=w_2=c$, $w_3=d^2$. It is easy to see that its
  solutions are the sequences of indices in $\{ 1^{i}.2^{i}.3^{i} \mid
  i\geq 1 \} \cup \{ 3^{i}. 2^{i}.1^{i} \mid i\geq 1 \}$.  In particular
  this language over the alphabet $\{1, 2, 3\}$ is not context-free and
  this will be useful in the sequel.  \nl Let now $\Ga $ be an alphabet
  having at least two elements such that $\Ga \cap \{c, d\} = \emptyset$,
  and $(u_1,\ldots ,u_n)$ and $(v_1,\ldots ,v_n)$ be two sequences of $n$
  non-empty words in $\Gas$. Let PCP be the Post Correspondence Problem of
  instances $(u_1,\ldots ,u_n)$ and $(v_1,\ldots ,v_n)$.

  Let $A=\{a, b\}$ and $C=\{c_1, \ldots , c_n\}$ and $D=\{d_1, d_2 , d_3\}$
  be three alphabets two by two disjoints. We assume also that $A \cap \{c,
  d\} = \emptyset$.

  We define the $\om$-rational function 
  $f$ of domain $\dom(f)=C^+.D^+.A^\om$ by : 
  \begin{itemize}
  \item $f(x)=u_{i_1}\ldots u_{i_k}.t_{j_1}\ldots t_{j_p}.z $~~if
    $x=c_{i_1}\ldots c_{i_k}.d_{j_1}\ldots d_{j_p}.z$ and $z\in
    (A^\star.a)^\om$.
  \item $f(x)=v_{i_1}\ldots v_{i_k}.w_{j_1}\ldots w_{j_p}.z $~~if
    $x=c_{i_1}\ldots c_{i_k}.d_{j_1}\ldots d_{j_p}.z$ and $z\in
    A^\star.b^\om$.
  \end{itemize}

  Reasoning as in the preceding proof and using the fact that
  $(A^\star.a)^\om$ and $A^\star.b^\om$ are both dense in $A^\om$, we can
  prove that the function $f$ is continuous at point $x=c_{i_1}\ldots
  c_{i_k}.d_{j_1}\ldots d_{j_p}.z$, where $z\in A^\om$, if and only if the
  sequence $i_1,\ldots ,i_k $ is a solution of the Post Correspondence
  Problem PCP of instances $(u_1,\ldots ,u_n)$ and $(v_1,\ldots ,v_n)$ and the
  sequence $j_1, \ldots , j_p$ is a solution of the Post Correspondence
  Problem PCP$_1$.  There are now two cases.

  \paragraph{First Case.} The Post Correspondence Problem PCP has not any
  solution. Thus the function $f$ has no points of continuity, i.e.
  $C(f)=\emptyset$.

  \paragraph{Second Case.}  The Post Correspondence Problem PCP has at
  least one solution $i_1,\ldots ,i_k $. We now prove that in that case the
  continuity set $C(f)$ is not a context-free $\om$-language, i.e. is not
  accepted by any B\"uchi pushdown automaton.  Towards a contradiction,
  assume on the contrary that $C(f)$ is a context-free $\om$-language.
  Consider now the intersection $C(f) \cap R$ where $R$ is the regular
  $\om$-language $c_{i_1}\ldots c_{i_k}.(d_1^+.d_2^+.d_3^+).A^\om$.  The
  class $CF_\om$ of context-free $\om$-languages is closed under
  intersection with regular $\om$-languages, \cite{sta}, thus the language
  $C(f) \cap R$ would be also context-free.  But $C(f) \cap R =
  c_{i_1}\ldots c_{i_k}.\{d_1^{i}.d_2^{i}.d_3^{i} \mid i\geq 1 \}.A^\om$
  and this $\om$-language is not context-free because the finitary language
  $\{d_1^{i}.d_2^{i}.d_3^{i} \mid i\geq 1 \}$ is not context-free.  So we
  have proved that $C(f)$ is not a context-free $\om$-language.

  In the first case $C(f)=\emptyset$ so $C(f)$ is a regular hence also
  context-free $\om$-language. In the second case $C(f)$ is not a
  context-free $\om$-language so it is not $\om$-regular.  But one cannot
  decide which case holds because one cannot decide whether the Post
  Correspondence Problem PCP has at least one solution $i_1,\ldots ,i_k $.
\end{proof}

\section{Continuity set of  synchronous $\om$-rational functions}

We have shown that for non-synchronous rational functions, the continuity
set can be very complex.  In this section, we show that the landscape is
quite different for synchronous rational functions.  Their continuity set is
always rational.  Furthermore we show that any ${\bf \Pi}_2^0$ rational set is 
the continuity set of some rational function.

\begin{thrm}
  Let $f : A^\omega \to B^\omega$ be a rational synchronous function.  The
  continuity set $C(f)$ of~$f$ is rational.
\end{thrm}
\begin{proof}
  We actually prove that the complement $A^\omega \setminus C(f)$ is
  rational.  Since the inclusion $C(f) \subseteq \dom(f)$ holds and the set
  $\dom(f)$ is rational, it suffices to prove that $\dom(f) \setminus C(f)$
  is rational. 

  Suppose that $f$ is realized by the synchronous transducer~$\mathcal{T}$. 
  Without loss of generality, it may be assumed that $\mathcal{T}$ is trim,
  that is, any state~$q$ appears in an accepting path.  Let $x$ be an
  element of the domain of~$f$.  We claim that $f$ is not continuous at~$x$
  if there are two infinite paths $\gamma$ and~$\gamma'$ in~$\mathcal{T}$
  such that the following properties hold,
  \begin{itemize}
  \item[i)] the path $\gamma$ is accepting and $y = f(x)$.
  \item[ii)] the labels of~$\gamma$ and~$\gamma'$ are $(x,y)$ and $(x,y')$
    with $y \neq y'$.
  \end{itemize}
  The path~$\gamma$ exists since $x$ belongs to the domain of~$f$.  Remark
  that the path~$\gamma'$ cannot be accepting since $\mathcal{T}$ realizes
  a function.  It is clear that if such a path~$\gamma'$ exists, the
  function~$f$ cannot be continuous at~$x$.  

  Suppose that $f$ is not continuous at~$x$.  There is a sequence
  $(x_n)_{n\ge0}$ of elements from the domain of~$f$ converging to~$x$ such
  that $d(f(x),f(x_n)) > 2^{-k}$ for some integer~$k$.  Since each~$x_n$
  belongs to the domain of~$f$, there is a path~$\gamma_n$ whose label is
  the pair $(x_n,f(x_n))$.  Since the set of infinite paths is a compact
  space, it can be extracted from the sequence $(x_n)_{n\ge0}$ another
  sequence $(x_{s(n)})_{n\ge0}$ such that the sequence
  $(\gamma_{s(n)})_{n\ge0}$ converges to a path~$\gamma'$.  Let $(x',y')$
  be the label of this path~$\gamma'$.  Since $(x_n)_{n\ge0}$ converges
  to~$x$, $x'$ is equal to~$x$ and since $d(f(x),f(x_n)) > 2^{-k}$, $y'$ is
  different from~$y$.  This proves the claim.

  From the claim, it is easy to build an automaton that accepts infinite
  words~$x$ such that $f$ is not continuous at~$x$.  Roughly speaking, the
  automaton checks whether there are two paths $\gamma$ and~$\gamma'$ as
  above.  Let $\mathcal{T}$ be the transducer $(Q,A,B,E,q_0,F)$.  We build
  a non deterministic B\"uchi automaton $\mathcal{A}$.  The state set of
  $\mathcal{A}$ is $Q \times Q \times \{0,1\}$.  The initial state is
  $(q_0,q_0,0)$ and the set of final states is $F \times Q \times \{1\}$.
  The set of transitions of this automaton is
  \begin{align*}
    G & = \{ (p,p',0) \trans{a} (q,q',0) \mid  \exists b\in B ~~
           p \trans{a|b} q \in E \text{ and } p' \trans{a|b} q' \in E\} \\
      & \cup \{ (p,p',0) \trans{a} (q,q',1) \mid  \exists b, b' \in B ~~ 
           p \trans{a|b} q \in E, p' \trans{a|b'} q' \in E \text{ and } b
           \neq b'\} \\
      & \cup \{ (p,p',1) \trans{a} (q,q',1) \mid \exists b, b' \in B ~~ 
           p \trans{a|b} q \in E \text{ and } p' \trans{a|b'} q' \in E\}
  \end{align*}
\end{proof}

\begin{thrm} \label{thm:charContSet}
  Let $X$ be a rational ${\bf \Pi}_2^0$ subset of~$A^\omega$.  Then $X$ is the
  continuity set $C(f)$ of some rational synchronous function~$f$ of
  domain~$A^\omega$.
\end{thrm}
\begin{proof}
  If $A$ only contains one symbol~$a$, the result is trivial since
  $A^\omega$ only contains the infinite word~$a^\omega$.  We now assume that $A$
  contains at least two symbols.  Let $b$ be a distinguished symbol in~$A$
  let $c$ a new symbol not belonging to~$A$.

  We  define a synchronous function~$f$ that is of the following
  form : 
  \begin{displaymath}
    f(x) = 
    \begin{cases}
      x & \text{if $x \in X$} \\
      wc^\omega \text{ for some prefix~$w$ of~$x$} & 
        \text{if $x \in \overline{X} \setminus X$} \\
      wb^\omega \text{ for some prefix~$w$ of~$x$} & 
        \text{if $x \in (A^*b)^\omega \setminus \overline{X}$} \\
      wc^\omega  \text{ for some prefix~$w$ of~$x$} & 
        \text{otherwise,}
    \end{cases}
  \end{displaymath}
\noindent where $w$ is a word precised below. 

  By Theorem~\ref{thm:landweber}, there is a deterministic B\"uchi
  automaton $\mathcal{A} = (Q,A,E,\{q_0\},F)$ accepting~$X$.  We assume
  that $\mathcal{A}$ is trim.
  
  The function~$f$ is now defined as follows.  If $x$ belongs to~$X$, then
  $f(x)$ is equal to~$x$.  If $x$ belongs to the closure $\overline{X}$ of~$X$ but not
  to~$X$, let $w$ be the longest prefix of~$x$ which is the label of a path
  in~$\mathcal{A}$ from~$q_0$ to a final state.  Then $f(x)$ is equal
  to~$wc^\omega$.  If $x$ does not belong to the closure of~$X$, let $w$ be
  the longest prefix which is the label of a path in~$\mathcal{A}$
  from~$q_0$.  Then $f(x)$ is equal to $wb^\omega$ if $b$ occurs infinitely
  many times in~$x$ and it is equal to $wc^\omega$ otherwise.  It is easy
  to verify that the continuity set of~$f$ is exactly~$X$.

  We now give a synchronous transducer~$\mathcal{T}$ realizing the
  function~$f$.  Let $R$ be the set of pairs $(q,a)$ such that there is no
  transition $q \trans{a} p$ in~$\mathcal{A}$.  The state set
  of~$\mathcal{T}$ is $Q \times \{0\} \cup (Q\setminus F) \times \{1\} \cup
  \{q_1,q_2,q_3,q_4\}$.  The initial state is $q_0$ and the set of final
  states is $F \times \{0\}\cup (Q\setminus F) \times \{1\} \cup
  \{q_2,q_4\}$.  The set of transitions is defined as follows.
 \begin{align*}
    G & = \{ (p,0) \trans{a|a} (q,0) \mid 
           p \trans{a} q \in E \} \\
      & \cup \{ (p,0) \trans{a|c} (q,1) \mid 
           p \trans{a} q \in E \text{ and } q \notin F \} \\
      & \cup \{ (p,1) \trans{a|c} (q,1) \mid 
           p \trans{a} q \in E \text{ and } p,q \notin F \} \\
      & \cup \{ (p,0) \trans{a|b} q_1 \mid 
            (p,a) \in R \} \\
      & \cup \{ (p,0) \trans{a|c} q_3 \mid 
            (p,a) \in R \} \\
      & \cup \{ p \trans{a|b} q_1 \mid 
           p \in \{q_1,q_2\} \text{ and } a \neq b\} \\
      & \cup \{ p \trans{b|b} q_2 \mid p \in \{q_1,q_2\}\} \\
      & \cup \{ q_3 \trans{a|c} q_3 \mid a \in A\} \\
      & \cup \{ p \trans{a|c} q_4 \mid 
           p \in \{q_3,q_4\} \text{ and } a \neq b\} 
  \end{align*}
\end{proof}
  
Recall that a point~$x$ of a subset~$D$ of a topological space~$X$ is
\emph{isolated} if there is a neighborhood of~$x$ whose intersection with~$D$
is equal to~$\{x\}$.  Isolated points have the following property with
regard to continuity.  Any function from a domain~$D$ is continuous at any
isolated point of~$D$.  Therefore, if $X$ is the continuity set of some
function of domain~$D$, $X$ must contain all isolated points of~$D$.  The
following theorem states that for rational ${\bf \Pi}_2^0$ sets, this condition
is also sufficient.

\begin{thrm}\label{X-D}
  Let $D$ and~$X$ be two rational subsets of~$A^\omega$ such that $X
  \subseteq D$.  If there exists a rational  ${\bf \Pi}_2^0$-subset $X'$ of $A^\omega$ such that 
$X=X' \cap D$,  and if $X$ contains all
  isolated points of~$D$, then it is the continuity set $C(f)$ of some
  synchronous rational function~$f$ of domain~$D$.
\end{thrm}

In the proof of Theorem~\ref{thm:charContSet}, the complement of the
set~$\overline{X}$ has been partitioned into two dense sets.  The following
lemma extends this result to any rational set of infinite words.

\begin{lmm} \label{lem:partition}
  Let $X$ be a rational set containing no isolated points.  Then, the
  set~$X$ can be partitioned into two rational sets $X'$ and~$X''$ such
  that both $X'$ and~$X''$ are dense in~$X$.
\end{lmm}
\begin{proof}
  Let $X$ be a rational set of infinite words with no isolated points and
  let $\mathcal{A}$ be a deterministic and trim Muller automaton
  accepting~$X$. We refer the reader for instance  to \cite{tho,pp} for the definition and properties of  Muller automata.
  From any state~$q$, either there is no accepting path
  starting in~$q$ or there are at least two accepting paths with different
  labels starting from~$q$.  

  We first consider the case where the table~$\mathcal{T}$ of~$\mathcal{A}$
  only contains one accepting set~$F$.  Since the automaton is trim, all
  states of~$F$ belong to the same strongly connected component
  of~$\mathcal{A}$.  We consider two cases depending on whether the set~$F$
  contains all states of its connected component.  

  Suppose first that  the state~$q$ does not belong to~$F$ but is in the same
  strongly connected component as~$F$.  Let $X'$ and~$X''$ the sets of
  words which respectively label an accepting path which goes an odd or an
  even number of times through the state~$q$.  It is clear that $X'$
  and~$X''$ have the required property.

  Suppose now that $F$ contains all the states in its strongly connected
  component.   Since $X$ has no isolated point, there must be a state~$q$
  of~$F$ with two outgoing edges $e$ and~$e'$.  Let $X'$ be the set of
  words which label a path of the following form.  The trace of this path
  over the two edges $e$ and $e'$ is an infinite sequence of the form
  $(e+e')^*(ee')^\omega$.

  We now come back to the general case where the table~$\mathcal{T}$
  of~$\mathcal{A}$ may contain several accepting sets $\{F_1,\ldots,F_n\}$.
  Let $X_i$ be the set of words accepted by the table $\{F_i\}$.  Note first
  that if the both sets $X_i$ and $X_j$ can be partitioned into dense
  rational sets into $X'_i$, $X''_i$, $X'_j$ and $X''_j$, the both sets
  $X'_i \cup X'_j$ and $X''_i \cup X''_j$ are dense in~$X_i \cup X_j$.

  Note also that if $F_i$ is accessible from~$F_j$, then any set dense in
  $X_i$ is also dense in~$X_j$ and therefore in~$X_i \cup X_j$.  It follows
  that if the set $X_i$ can be partitioned into two dense rational sets
  $X'_i$ and $X''_i$, the set $X_i \cup X_j$ can be partitioned into $X'_i
  \cup X_j$ and $X''_i$.

  From the previous two remarks, it suffices to partition independently
  each~$X_i$ such that the corresponding~$F_i$ is maximal for
  accessibility.  By maximal, we mean that $F_i$ is maximal whenever if
  $F_j$ is accessible from~$F_i$, then $F_i$ is also accessible from~$F_j$
  and both sets $F_i$ and~$F_j$ are in the same strongly connected
  component.  This can be done using the method described above.
\end{proof}

We now come to the proof of the previous theorem.
\begin{proof}
  The proof is similar to the proof of Theorem~\ref{thm:charContSet} but
  the domain~$D$ has to be taken into account.  By hypothesis there exists 
 a  rational ${\bf \Pi}_2^0$-subset $X'$ of $A^\omega$ such
  that $X = X' \cap D$.  Let $\mathcal{A}$ a trim and deterministic B\"uchi
  automaton accepting~$X'$.

  We define the function~$f$ of domain~$D$ as follows.  For any $x$ in~$X$,
  $f(x)$ is still equal to~$x$.  If $x$ belongs to $\overline{X} \cap D
  \setminus X$, $f(x)$ is equal to~$wc^\omega$ where $w$ is the longest
  prefix of~$x$ which is the label in~$\mathcal{A}$ of a path form the
  initial state to a final state. 

  By Lemma~\ref{lem:partition}, the set $Z = D \setminus \overline{X}$ can
  be partitioned into two rational subsets $Z_1$ and $Z_2$ such that both
  $Z_1$ and~$Z_2$ are dense into~$Z$.  If $x$ belongs to~$D \setminus
  \overline{X}$, then $f(x)$ is defined as follows.  Let $w$ be the longest
  prefix of~$x$ which is the label in~$\mathcal{A}$ of a path from the
  initial state.  Then $f(x)$ is equal to $wb^\omega$ if $x \in Z_1$ and
  $f(x) = wc^\omega$ if $x \in Z_2$.
\end{proof}

\noi The following corollary provides a complete characterization of sets of continuity of synchronous 
rational functions of domain~$D \subseteq A^\omega$ when 
$D$ is the intersection of a rational ${\bf \Si}_2^0$-subset  and of a  ${\bf \Pi}_2^0$-subset  of $A^\omega$. 
This is in particular the case if $D$ is simply  a 
${\bf \Si}_2^0$-subset  or a  ${\bf \Pi}_2^0$-subset  of $A^\omega$.

\begin{crllr}\label{cor}
Let $D$ and~$X$ be two rational subsets of~$A^\omega$ such that $X$ is a ${\bf \Pi}_2^0$-subset  of 
 $D$ and $D=Y_1 \cap Y_2$, where  $Y_1$ is a rational  
${\bf \Si}_2^0$-subset  of $A^\omega$ and  $Y_2$ is a ${\bf \Pi}_2^0$-subset  of $A^\omega$. 
If $X$ contains all
  isolated points of~$D$, then it is the continuity set $C(f)$ of some
  synchronous rational function~$f$ of domain~$D$.
\end{crllr}

\begin{proof}
Let $D$ and~$X$ satisfying the hypotheses of the corollary.  

 Assume firstly that $D=Y_1$ is a ${\bf \Si}_2^0$-subset  of $A^\omega$. 
Then it is easy to see that 
$X'=X \cup ( A^\omega - D)$ is a rational ${\bf \Pi}_2^0$-subset  of $A^\omega$ such that $X=X' \cap D$. 
Thus in this case  Corollary \ref{cor} follows from Theorem \ref{X-D}.

Assume now that $D=Y_1\cap Y_2$, where $Y_1$ is a rational  
${\bf \Si}_2^0$-subset  of $A^\omega$ and  $Y_2$ is a ${\bf \Pi}_2^0$-subset  of $A^\omega$. 
By hypothesis $X$ is a ${\bf \Pi}_2^0$-subset of $D$ thus there is a ${\bf \Pi}_2^0$-subset $X_1$ of $A^\omega$ such that 
$X = X_1\cap D = X_1 \cap (Y_1\cap Y_2)  = (X_1 \cap Y_2)  \cap Y_1$. This implies that $X$ is also a  ${\bf \Pi}_2^0$-subset of
$Y_1$ because $X_1 \cap Y_2$ is a ${\bf \Pi}_2^0$-subset  of $A^\omega$ as intersection of two ${\bf \Pi}_2^0$-subsets  of $A^\omega$. 
From the first case we can infer that  there is a rational ${\bf \Pi}_2^0$-subset $X'$ of $A^\omega$ such that $X=X' \cap Y_1$. 
Now we have also $X = X' \cap (Y_1 \cap Y_2) = X' \cap D$ because $X \subseteq Y_2$. 
Thus in this case  again Corollary \ref{cor} follows from Theorem \ref{X-D}.
\end{proof}

\begin{footnotesize}

\end{footnotesize}

\begin{thebibliography}{99}





\bibitem[BT70]{bt} Ya M. Barzdin and B.A. Trakhtenbrot, Finite Automata, 
Behaviour and Synthesis, Nauka, Moscow, 1970 (English translation, 
North Holland, Amsterdam, 1973).

\bibitem[BC00]{bc} M.-P. Béal and O. Carton, 
Determinization of Transducers over Infinite Words, in the Proceedings of the 
International Conference ICALP 2000, U. Montanari et al., eds.,  
Lecture Notes in Computer Science, Volume 1853,  2000, p. 561-570. 

\bibitem[BCPS00]{bcps} M.-P. B\'eal ,  O. Carton, C.  Prieur and  J. Sakarovitch,  
Squaring Transducers: An Efficient Procedure for Deciding Functionality 
and Sequentiality, 
Theoretical Computer Science, Volume 292 (1), p. 45-63, 2003. 


\bibitem[Ber79]{ber} J. Berstel, Transductions and Context Free Languages, 
Teubner Verlag, 1979. 


\bibitem[B\"uc62]{bu62}  J.R. B\"uchi, On a Decision Method in Restricted 
Second Order Arithmetic, 
Logic Methodology and Philosophy of Science, Proc. 1960 Int. Congr., Stanford
 University Press, 1962, p. 1-11.



\bibitem[Cho77]{cho77}  C. Choffrut, 
Une Caract\'erisation des Fonctions S\'equentielles et 
des Fonctions Sous-S\'equentielles en tant que Relations Rationnelles, 
Theoretical  Computer Science, Volume 5,  1977, p.325-338. 


\bibitem[CG99]{cg} C. Choffrut and  S. Grigorieff, Uniformization of Rational Relations, 
Jewels are Forever 1999,  
J. Karhumäki, H. Maurer, G. Paun and G. Rozenberg editors, Springer,  p.59-71. 


\bibitem[EH93]{eh} J. Engelfriet and H. J. Hoogeboom, X-automata on $\om$-Words, 
Theoretical  Computer Science, Volume 110 (1),  1993,  p.1-51.


\bibitem[Fin03a]{relrat} O. Finkel, On the Topological 
Complexity of Infinitary Rational Relations, 
RAIRO-Theoretical Informatics and Applications, Volume 37 (2),  2003,  p. 105-113.

\bibitem[Fin03b]{relratdec} O. Finkel, Undecidability of Topological and Arithmetical 
Properties  of Infinitary Rational Relations, 
RAIRO-Theoretical Informatics and Applications, Volume 37 (2),  2003, p. 115-126.


\bibitem[Fin03c]{relratbor} O. Finkel, On 
 Infinitary Rational Relations and Borel Sets, 
in the Proceedings of the 
Fourth International Conference on Discrete Mathematics and Theoretical 
Computer Science DMTCS'03, 
7 - 12 July 2003, Dijon, France, Lecture Notes in Computer Science, Springer, 
Volume 2731,  p. 155-167. 


\bibitem[Fin06a]{stacs06} O. Finkel, On 
the Accepting Power of 2-Tape B\"uchi Automata,  
in the Proceedings of the  23rd International Symposium on Theoretical Aspects of Computer Science,
STACS 2006, Marseille, France, February 23-25, 2006, Lecture Notes in Computer Science, Volume 3884,  Springer, 
Volume 3884,  p. 301-312.

\bibitem[Fin06b]{mscs06} O. Finkel, 
Borel Ranks and Wadge Degrees of Omega Context Free Languages, 
Mathematical Structures in Computer Science, Volume 16 (5),  2006, p. 813-840.


\bibitem[FS93]{fs} C. Frougny and J. Sakarovitch, Synchronized rational Relations of Finite and Infinite Words, 
Theoretical Computer Science, Volume 108, 1993,  p. 45-82. 

\bibitem[Gir81]{gire1} F. Gire, Relations Rationnelles Infinitaires, Th\`{e}se de 
troisi\`{e}me cycle, Universit\'e Paris 7, Septembre 1981. 

\bibitem[Gir83]{gire} F. Gire, Une Extension aux Mots Infinis de la Notion de 
Transduction Rationnelle, 
6th GI Conf., Lecture Notes in Computer Science,  Volume 145, 1983, p. 123-139. 

\bibitem[GN84]{gn} F. Gire and M. Nivat,  Relations Rationnelles Infinitaires, 
Calcolo,  Volume XXI,  1984,  p. 91-125.  


\bibitem[Kec95]{kec} A.S.  Kechris, Classical Descriptive Set Theory, Springer-Verlag, 1995. 

\bibitem[Kur66]{ku} K. Kuratowski, Topology, Academic Press, New York 1966.

\bibitem[KMS89]{kms}  A. S. Kechris,  D. Marker, and R. L. Sami, $\Pi_1^1$ Borel Sets, 
The Journal of Symbolic Logic, Volume 54 (3), 1989, p. 915-920. 

\bibitem[Lan69]{la} L. H. Landweber, Decision Problems for $\om$-Automata, 
Math. Syst. Theory,  Volume 3 (4), 1969,  p. 376-384.


\bibitem[LT94]{lt} H. Lescow and W. Thomas, Logical Specifications of Infinite Computations,
 in: A Decade of Concurrency,  J. W. de Bakker et al., eds., 
Lecture Notes in Computer Science, Springer, Volume  803, 1994,  p. 583-621.

\bibitem[LS77]{ls} R. Lindner and L. Staiger, 
Algebraische Codierungstheorie - Theorie der Sequentiellen Codierungen, Akademie-Verlag, 
Berlin, 1977. 

\bibitem[Mos80]{mos} Y. N. Moschovakis, Descriptive Set Theory, North-Holland, Amsterdam 1980.

\bibitem[PP04]{pp} D. Perrin and J.-E. Pin, Infinite Words, Automata, Semigroups, Logic and Games, Volume 141 of 
Pure and Applied Mathematics, Elsevier, 2004. 

\bibitem[Pin96]{pin} J-E. Pin, Logic, Semigroups and Automata on Words, 
Annals of Mathematics and 
Artificial Intelligence,  Volume  16 (1996), p. 343-384.

\bibitem[Pri00]{pri} C. Prieur, Fonctions Rationnelles de Mots Infinis et Continuit\'e, 
Th\`{e}se de Doctorat, Universit\'e Paris 7, Octobre 2000. 

\bibitem[Pri01]{pri01} 
C. Prieur,  How to Decide Continuity of Rational Functions on Infinite Words, 
Theoretical Computer Science, Volume   250  (1-2),  2001,  p. 71-82. 


\bibitem[Sim92]{sim} P. Simonnet, Automates et Th\'eorie Descriptive, 
Th\`{e}se de Doctorat, Universit\'e 
Paris 7, March 1992.

\bibitem[Sta86]{stac} L. Staiger, Hierarchies of Recursive $\om$-Languages, Jour. Inform. 
Process. Cybernetics  EIK 22 ,1986, 5/6,  p. 219-241.


\bibitem[Sta97]{sta} L. Staiger, $\om$-Languages, 
Chapter of the Handbook of Formal languages,  Volume 3,
 edited by G. Rozenberg and A. Salomaa, Springer-Verlag, Berlin.

\bibitem[SW78]{sw} L. Staiger and K. Wagner, 
Rekursive Folgenmengen I, Z. Math Logik Grundlag. 
Math.,  Volume  24, 1978,  p. 523-538. 

\bibitem[Tho89]{tho89}  W. Thomas, Automata and Quantifier Hierarchies,  in: Formal Properties 
of Finite automata and Applications, Ramatuelle, 1988, Lecture Notes in Computer Science,  Volume  
386, Springer, Berlin, 1989, p.104-119. 


\bibitem[Tho90]{tho}  W. Thomas, Automata on Infinite Objects, in: J. Van Leeuwen, ed.,
 Handbook of Theoretical Computer Science, Volume B, Elsevier, Amsterdam, 1990, p. 133-191.


\end{thebibliography}
\end{document}